**Long Range Stress Correlations in the Inherent Structures of Liquids at Rest**

Sadrul Chowdhury, Sneha Abraham, Toby Hudson and Peter Harrowell

*School of Chemistry, University of Sydney, Sydney 2006 NSW, Australia*

Abstract

Simulation studies of the atomic shear stress in the local potential energy minima (inherent structures) are reported for binary liquid mixtures in 2D and 3D. These inherent structure stresses are fundamental to slow stress relaxation and high viscosity in supercooled liquids. We find that the atomic shear stress in the inherent structures (IS) of both liquids at rest exhibits slowly decaying anisotropic correlations. We show that the stress correlations contributes significantly to the variance of the total shear stress of the IS configurations and consider the origins of the anisotropy and spatial extent of the stress correlations.

## 1. INTRODUCTION

Microscopic accounts of the physical properties of low temperature liquids rely heavily on the use of local minima in the potential energy surface of the configuration space made up of the coordinates of all N particles [1]. These local minima, referred to as inherent structures (IS), represent local groundstates for the liquid. As the temperature is decreased, the trajectory of the liquid through its configuration space exhibits an increasing tendency to get trapped about local minima until eventually escaping via an activated process. This means that the properties of the liquid will increasingly reflect those of the associated inherent structure as the temperature is lowered. The idea of treating a liquid in terms of reference solids represents an explicit realization of the view that goes back to Frenkel's 1946 proposition [2] that rigidity coexists with the fluidity of a liquid so that as the latter property decreases, on cooling, the features of the underlying solid reference states become manifest.

Inherent structures have proven particularly useful in treating the (metastable) equilibrium properties of the supercooled liquid. The configurational entropy has been obtained from the dispersion of inherent structures in energy [3] and the equation of state of the supercooled liquid has been expressed as the ensemble of inherent structures treated as Einstein solids [4]. The relationship between the inherent structures and relaxation kinetics is less direct. Detailed analysis [5] of the slow relaxation dynamics has demonstrated that multiple transitions can occur between a set of IS's before the trajectory irreversibly escapes from its initial region of configuration space. It is this collection of IS's, referred to as a metabasin [6], rather than the individual minima, that represent the persistent physical constraint that requires the activated transitions for their relaxation.

The fundamental measure of 'fluidity' or its inverse, shear viscosity, is directly related to the relaxation of the shear stress fluctuations. The enormous increase in the shear viscosity as the glass transition is approach on cooling is a direct consequence of the growing persistence of shear stress fluctuations [7]. Recently, we demonstrated [8] that the inherent structures exhibited a distribution of shear stress. This residual or IS stress represents an obvious candidate for the persistent stress fluctuations at the heart of viscoleastic behaviour and, ultimately, the glass transition. Since the slow component of the shear stress is really that associated with a metabasin rather than an individual inherent structure, the correspondence will not be exact. Nevertheless, the IS stress provides an accessible nontrivial case of persistent stress fluctuations in amorphous materials and so their characterization is of interest.

In this paper we shall examine the IS stress fluctuations in 2D and 3D glass forming mixtures. We shall establish that long range correlations in stress exist in both 2D and 3D inherent structures and examine the how the fluctuations in the atomic stress contribute to the total variance of the inherent structure stress. The origin of the anisotropy of the stress correlation is examined and the significance of the inherent structure stress with respect to the mechanical properties of the supercooled liquid is discussed.

## 2. MODEL AND ALGORITHM

We have carried out constant NVT molecular dynamics (MD) simulations of two well studied models of glass forming binary alloys. In 2D, we have examined a binary mixture of soft disks introduced by Perera and Harrowell [9] and, in 3D, a binary Lennard-Jones (LJ) mixture due to Wahnström [10]. These potentials have been truncated at a distance $r_c = 4.5 a_{22}$ where $a_{22}$ is the larger diameter and the potential shifted so that $\tilde{\phi}_{ij}(r) = \phi_{ij}(r) - \phi_{ij}(r_c)$. The following reduced units are used throughout the paper: length =length/$a_{11}$, time $t = t/(m_1 a_{11}^2 / \varepsilon)^{1/2}$, temperature $T = k_B T / \varepsilon$, pressure (stress) $P = P a_{11}^3 / \varepsilon$ and energy $E = E / \varepsilon$. Unless otherwise indicated, the following number density $\rho$ and number of particles N were used: $\rho$ =0.7468 and N=1024 (2D) and $\rho$ =0.75 and N=1024 (3D). At these densities, the freezing temperatures for the respective crystal phases are $T_f(3D) = 0.6$ [7] and $T_f(2D) \sim 0.70$ [11]. Inherent structures are generated by carrying out a conjugate gradient minimization of the potential energy from configurations generated at time intervals of dt = 0.003 $\tau_\alpha$ along a given trajectory, where $\tau_\alpha$ is the stress relaxation time obtained from a stretched exponential fit to the long time tail of the shear stress autocorrelation function. For the calculations of the spatial correlations of stress presented in the following Sections, we have averaged over 20,000 and 5,000 inherent structures for the 2D and 3D liquids, respectively.

## 3.1 LONG RANGE CORRELATIONS OF THE INHERENT STRUCTURE STRESS

The total shear stress of an inherent structure $\sigma_{xy}$ can be written in terms of the sum over the contributions per particle $s_{xy}(i)$ , i.e.

$$\sigma_{xy} = \frac{1}{V} \sum_{i=1}^{N} s_{xy}(i) \tag{1}$$

where

$$s_{ab}(i) = -\frac{1}{2} \sum_{j \neq i}^{N} \frac{1}{|r_{ij}|} \frac{d\phi}{d|r_{ij}|} r_{ij}^{a} r_{ij}^{b} \tag{2}$$

where $a$ and $b$ refer to choices of the Cartesian axes and $r^{a}$ is the projection of the vector $\vec{r}$ onto the $a$ axes. In the following discussion we shall refer to $s_{ab}(i)$ as the atomic stress even through, strictly, we should divide this quantity by a local volume in order for it to have units of stress. The choice of local volume plays no role in the contribution of the atomic stress to the total stress as defined in Eq. 1.

The total IS shear stress variance $< \sigma_{xy}^{2} >$ can be expressed in terms of the atomic shear stress $s_{xy}$ as follows,

$$< \sigma_{xy}^{2} > = \frac{1}{V^2} \left( \sum_{i}^{N} \sum_{j}^{N} < s_{xy}(i) s_{xy}(j) > \right) \tag{3}$$

and the relevant atomic shear stress correlation function $C(\vec{r})$ is defined as

$$C(\vec{r}) = \frac{1}{N} \sum_{i} \sum_{j} < s_{xy}(i) s_{xy}(j) > \delta(\vec{r} - \vec{r}_{ij}) . \tag{4}$$

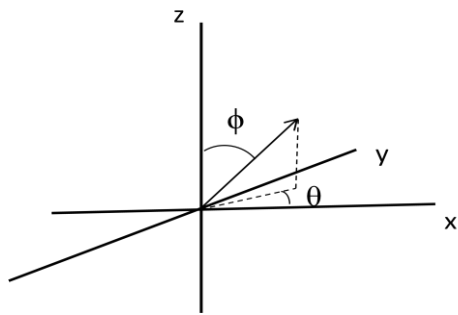

**Figure 1.** A diagram identifying the angular components θ and ϕ of the vector $\vec{r}$ .

To depict the shear stress correlation function in the 3D liquid we shall make use of a number of projections. As shown in Fig. 1, the xy plane includes the angle θ while the angle ϕ corresponds to directions out of the xy plane with ϕ = π/2 corresponding to the xy plane . In Fig. 2 we plot C(r,θ,ϕ) in the (θ,ϕ) plane at a radial distance of r = 1.0. The striking peaks encountered as we vary θ at ϕ = π/2 will be addressed shortly. The main message from Fig. 2 is the stress correlations decay quickly in the ϕ direction to a value that shows little variation in θ. We conclude that in the 3D liquid, the shear stress correlations are dominated by 2D like correlations in the plane corresponding to the Cartesian coordinates used to define the shear stress.

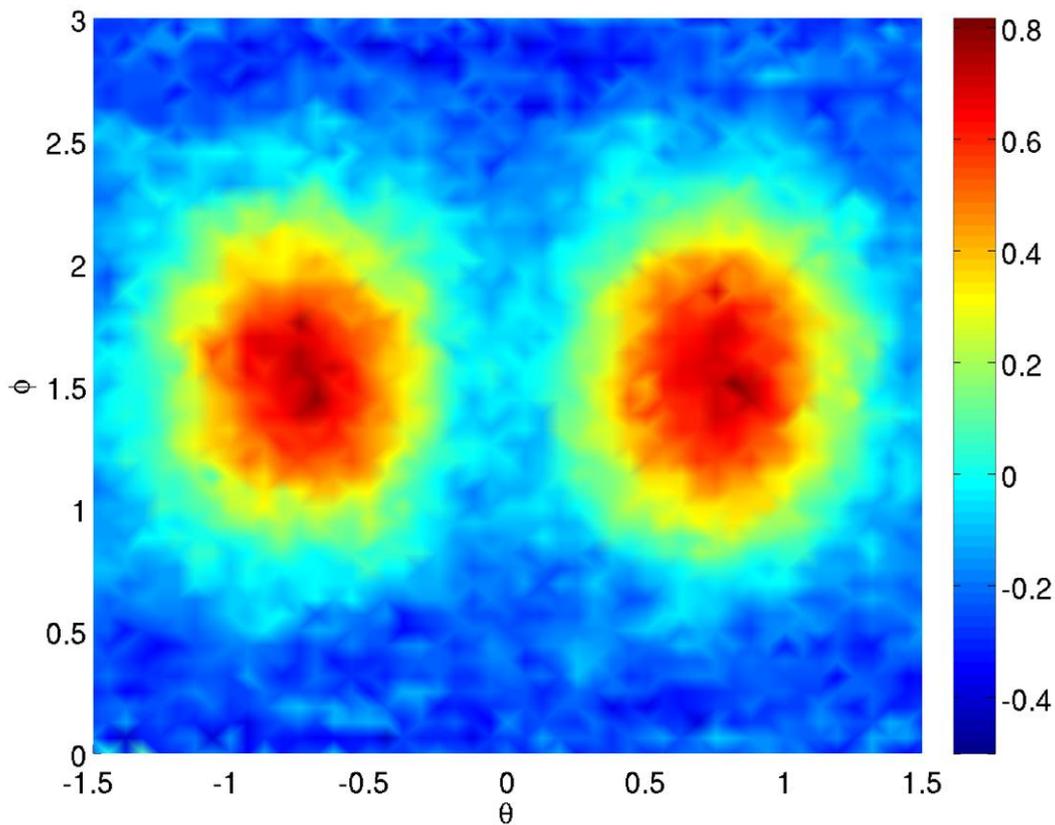

**Figure 2.** $C(r,\theta,\phi)$ for the 3D (L-J) mixture plotted as a function of $(\theta,\phi)$ for r = 1.0 for the inherent structures of the 3D mixture generated from a parent liquid at $T_{pl}$=0.60. The magnitude of $C(\bar{r})$ is color coded, as indicated, with red positive and blue negative.

In Fig. 3 we plot the projection of $C(\bar{r})$ for the 3D LJ mixture onto the xy plane. We find the shear stress correlations in the xy plane to be both anisotropic and slowly decaying. The anisotropy of the stress correlations in the xy plane are characterised by a $\cos 4\theta$ symmetry with the positive correlation aligned along the diagonals of the xy plane of the simulation cell. The decay of amplitude of $C(\bar{r})$ with respect to r along the diagonal is reasonably modelled as $1/r^2$ (see Fig. 4). In the 2D mixture, the atomic shear stress correlations exhibit the same anisotropy and decay with respect to r (see Fig. 4 and 5).

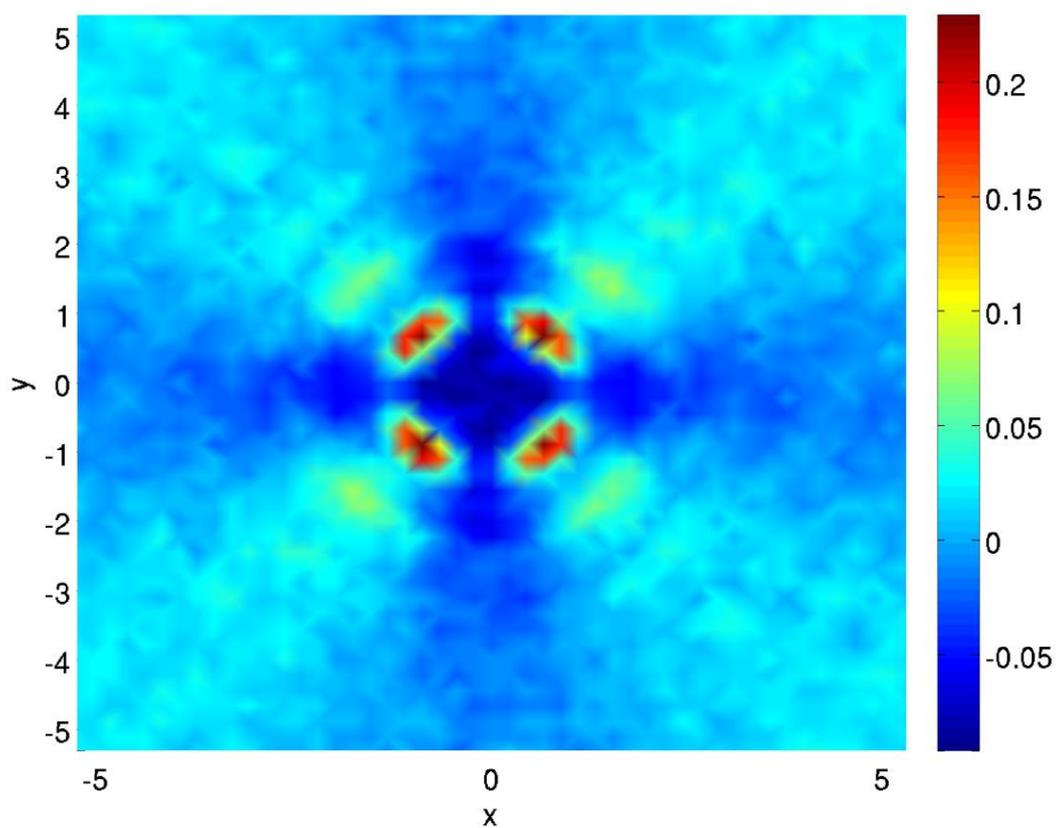

**Figure 3.** The plot of the atomic shear stress correlation function $C(\vec{r})$ projected onto the xy plane for the inherent structures of the 3D mixture generated from a parent liquid at $T_{pl}$=0.60. The conditions and plotting conventions are the same as Fig. 2.

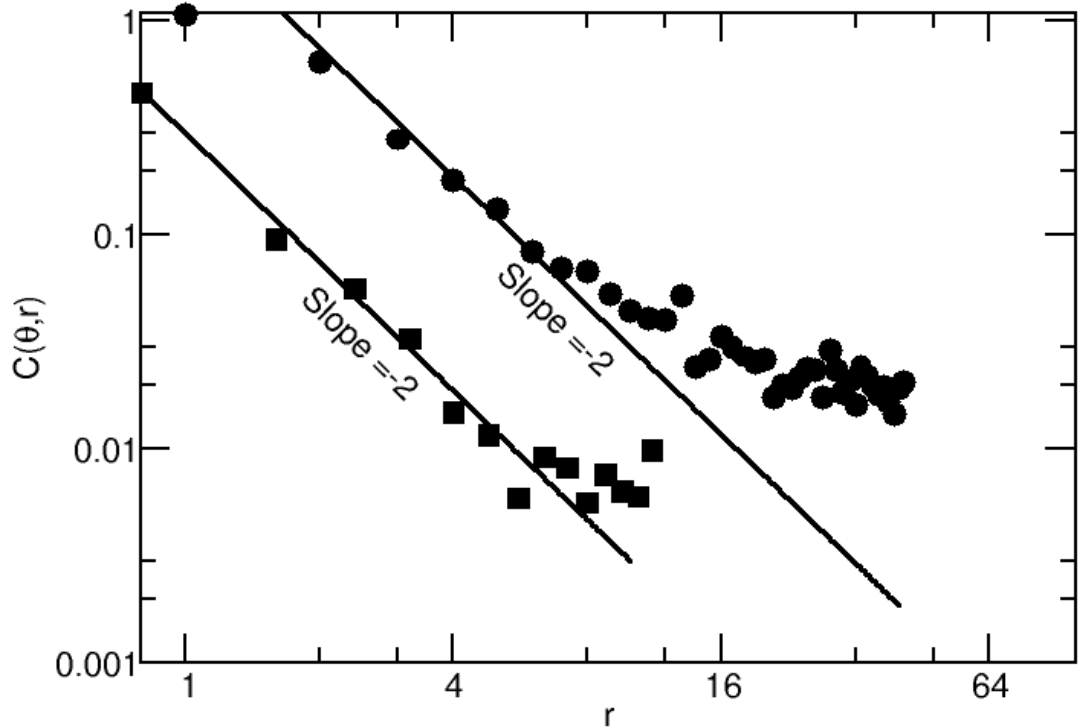

**Figure 4.** A log-log plot of the decay of $C(\theta, r)$ with respect to r for $\theta = n\frac{\pi}{4}$ in the 2D liquid inherent structures with $T_{pl} = 0.40$ (circles) and in the 3D liquid inherent structures with $T_{pl} = 0.60$ (squares). Note that the $1/r^2$ decay for the 3D liquid refers specifically to the decay of $C(\vec{r})$ after being projected onto the xy plane. Note the $1/r^2$ behaviour extends up to ~5 particle diameters beyond which we see deviations that we associate with periodic boundary effects.

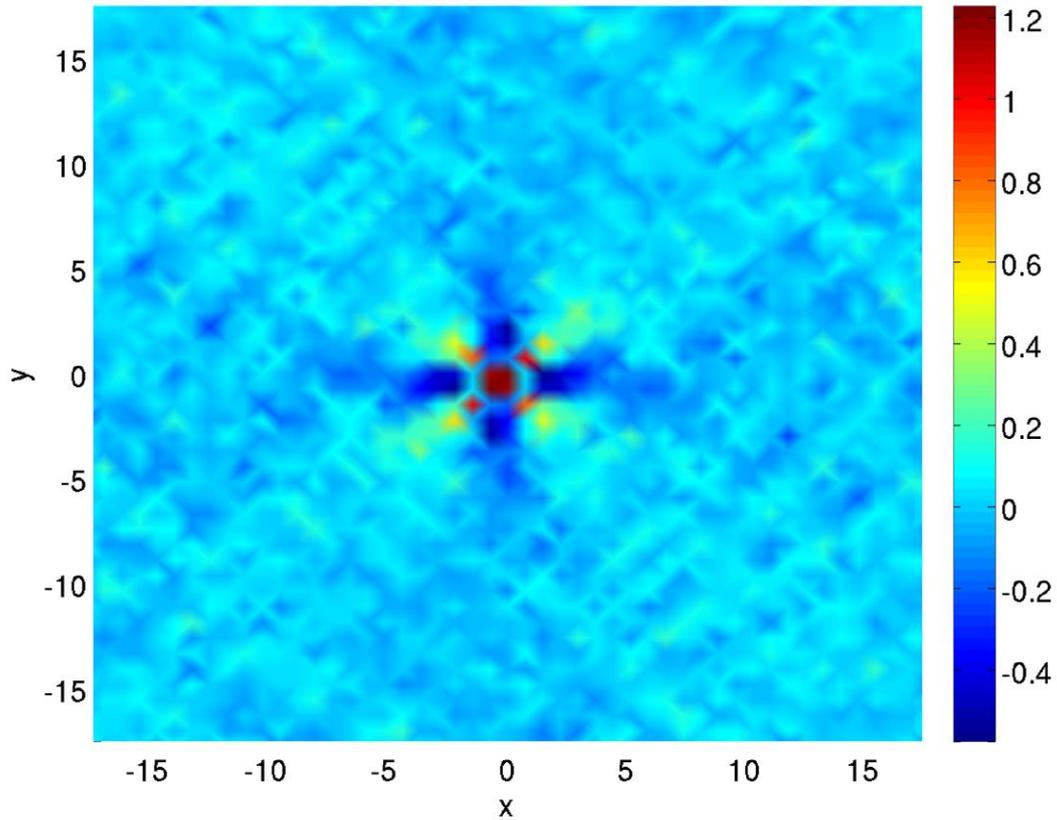

**Figure 5.** The plot of the atomic shear stress correlation function $C(\vec{r})$ as defined in Eq. 4 for the inherent structures of the 2D mixture generated from a parent liquid at $T_{pl} = 0.365$. The 2D plot has been obtained by averaging over a thin slab in the vertical direction.

## 3.2 CONTRIBUTIONS TO THE STRESS VARIANCE

Previously [8], we established that the variance of the IS stress varied with system size as $1/N$. This result is reminiscent of the central limit theorem result that would apply to the case where we are summing $N$ uncorrelated atomic shear stresses. However, having just demonstrated that the shear stress exhibits a power law dependence, we are dealing with anything but uncorrelated stress and, hence, the origin of the size dependence of $< \sigma_{xy}^2 >$ represents a puzzle.

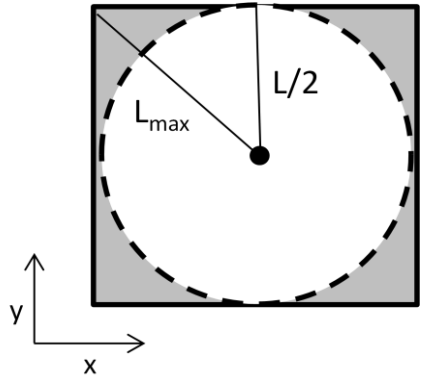

**Figure 6.** A sketch of the integration domains referred to in Eq. 5. The small filled circle represents the self term, the white dotted disk depicts the second term, that over which full angular averaging is possible and the shaded remainder of the cell corresponds to the third term in Eq. 5.

To better understand the relationship between the spatial correlation is stress and the scaling with N, we shall resolve $< \sigma_{xy}^2 >$ into three components (as depicted in Fig. 6),

$$< \sigma_{xy}^2 >= \frac{1}{V^2} \left[ N < s_{xy}^2 > + \int\limits_{1}^{L/2} d\vec{r}\,\overline{n}(\vec{r})C(\vec{r}) + \int\limits_{L/2}^{L\max} d\vec{r}\,\overline{n}(\vec{r})C(\vec{r}) \right] \qquad (5)$$

where L is the length of the simulation cell, $L_{max}$ is the distance from the centre to a vertex of the cell and $\overline{n}(\vec{r}) = < \sum_{i}\sum_{j}\delta(\vec{r}-\vec{r}_{ij}) >$. The first term in Eq. 5 is the contribution of the individual atomic stresses, the 'self' term. The distribution of atomic shear stress for the inherent structures of the 2D and 3D liquids are found to be zero mean Gaussians with variances $< s_{xy}^2 >$ that display only a weak dependence on $T_{pl}$ as shown in Fig. 7. The second term in Eq. 5 is the contribution from the local correlations in the atomic stress. In Fig. 8 we

plot the radially averaged stress correlation function out to a distance L/2 for a number of

different system sizes. We note that the stress is anti-correlated with the nearest neighbours

and the angular averages stress correlations decay within 3-4 atomic diameters. The short

range of this correlation, relative to that shown in Fig. 3, reflects of the angular averaging

over the positive and negative components of the angular distribution.  We find no significant

dependence of this decay length on $T_{pl}$. For values of r > L/2, where L is the length of the

simulation box, this angular averaging can no longer be carried out. Due to the square shape

of the simulation cell, the angular averaging over these distances will be restricted to the cell

'corners'. We have separated out this contribution to $< \sigma_{xy}^2 >$ as the third term in Eq. 5

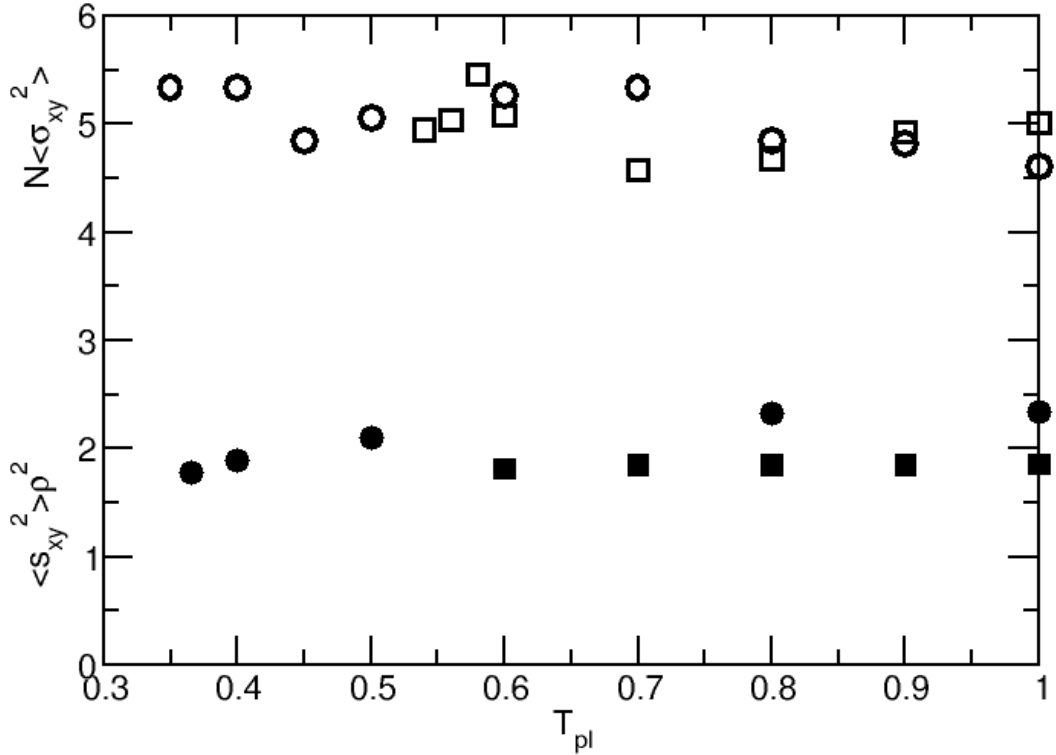

**Figure 7.** The variance of total shear stress (open symbols) and atomic shear stress (filled symbols) in the 2D (circle) and 3D (square) inherent structures as a function of T$_{pl}$, the parent liquid temperature.

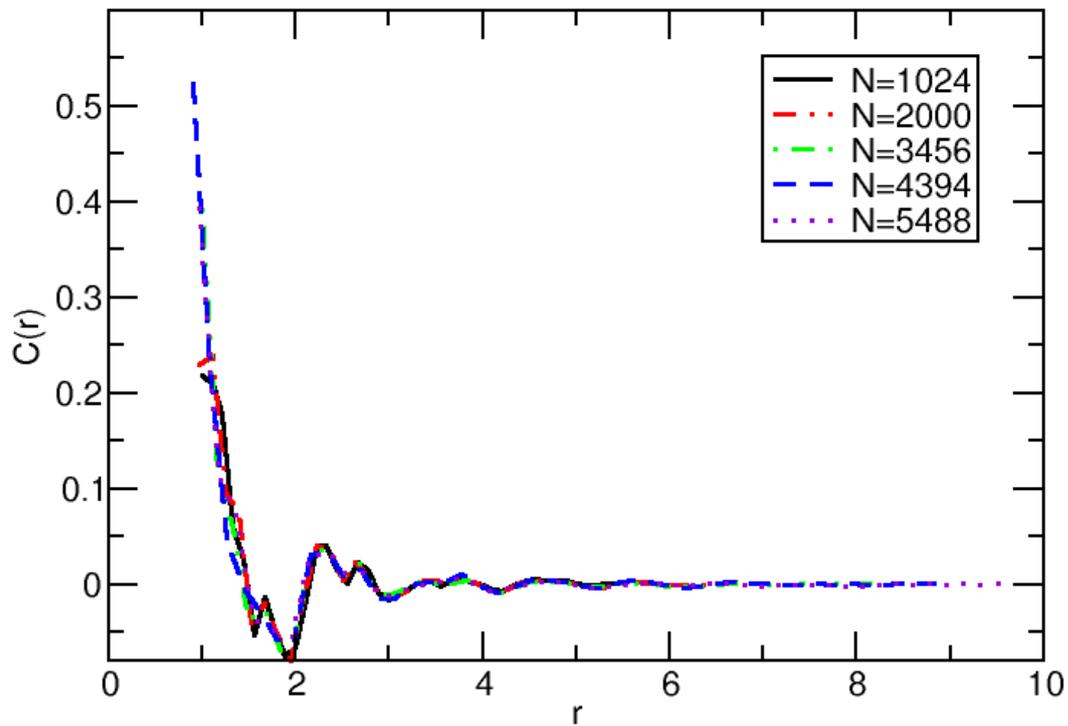

**Figure 8.** The angular averaged atomic shear stress correlation function C(r) for the inherent structures of the 3D liquid at T = 0.60. The results for a number of different system sizes are shown.

In Fig.9 we present the value of $N < \sigma_{xy}^2 >$ resolved into these three components for a range of system sizes. We find that they all contribute significantly to the overall magnitude of shear stress fluctuations of the inherent structures, with no individual contribution showing any significant dependence on system size N, i.e. each individual contribution to $< \sigma_{xy}^2 >$ scales as $\sim 1/N$.

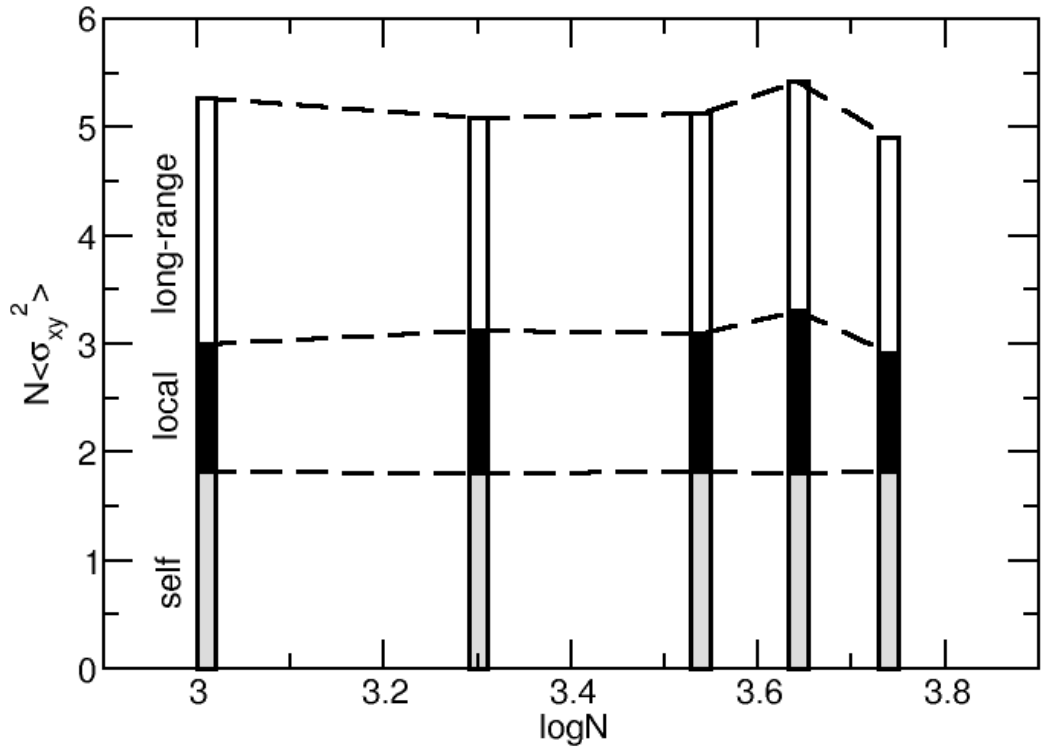

**Figure 9.** The contributions to the overall shear stress variance of the three components

identified in Eq. 5; i.e. the self (1st term), local (2nd term) and long-range (3rd term) for 3D

liquids of different size with $T_{pl} = 0.60$. The dashed lines are only to guide the eye.

## 3.3 THE ORIGIN OF THE ANISOTROPY OF THE SHEAR STRESS

## CORRELATIONS

Previous studies [12-16] of the strain correlations in liquids sheared in the nonlinear regime have reported on the extended strain fields associated with local reorganization events, correlations related to Eshelby's [17] calculations of local distortions in an elastic continuum. In this paper we have established that anisotropic stress correlations are a property of a *quiescent* liquid. A similar observation has been recently reported by Wu *et al* [18]. Chattoraj and Lemaître [19] and Lemaître [20] have reported the presence of analogous anisotropies in the non-affine strain in a liquid undergoing Newtonian (i.e. linear) shear flow. Since linear behaviour implies that the response is that of the equilibrium liquid, it makes sense that the $\cos(4\theta)/r^2$ strain correlation observed under the linear shear flow is actually the correlation of the equilibrium liquid, as shown here. In refs. [18-20] the authors argue that the anisotropic strain correlations are the results of accumulated Eshelby-like rearrangements. We shall argue here that this is not the case.

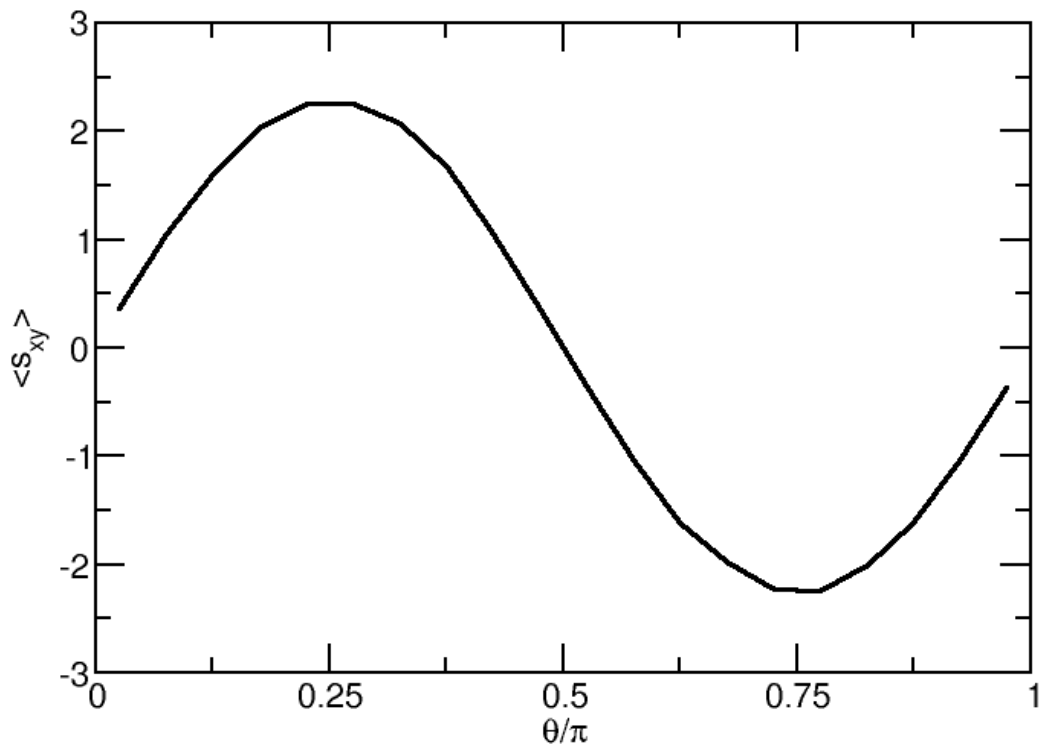

**Figure 10.** The average value of the atomic shear stress $s_{xy}$ as a function of the angle $\theta$ between the principal stress axes and the x axes of the inherent structures of the 2D liquid with $T_{pl}$= 0.40. The points of maximum magnitude correspond to the diagonals $\pi/4$ and $3\pi/4$. The curve is well represent by $\sin \theta$ supporting the assertion that the principle stress axes is randomly oriented.

If the observed angular correlation is not the result of an accumulation of Eshelby events, where does it come from? A complete description of the stress in the plane on an individual atom is provided by the 2x2 stress tensor. The difference between the eigenvalues of this tensor correspond to the maximum value $\hat{s}$ of the shear stress one can obtain for that atom by rotation of coordinates. (The constant volume constraint introduces a constraint on the trace of the total stress tensor. We have assumed here that the impact of this global constraint on the individual atomic stress tensor is negligible.) The principle stress axes for each atom $\vec{u}$ is defined as the eigenvector associated with the maximum eigenvalue. If we assume that the magnitude of $\hat{s}$ is independent of the orientation of the principal stress axis, then it would follow that the maximum magnitude of $<s_{xy}>$ will correspond with those cases where an atom's principal stress axis $\vec{u}$ is aligned along the positive and negative xy diagonals. To test this we plot $<s_{xy}>$ against $\theta$, where $\vec{u} \cdot \vec{x} = \cos\theta$, in Fig. 10. We find that the maximum amplitude of $<s_{xy}>$ does indeed occur at the diagonals, $\theta = \pi/4$, $3\pi/4$. Furthermore, the dependence of $<s_{xy}>$ on the orientation of the principal stress axis is well described by $\sin\theta$ indicates that $\vec{u}$ is *randomly* oriented. We conclude that, rather than the accumulation of flow induced Eshelby events, the anisotropy of the stress correlations is a simple geometrical consequence of the use of an arbitrary lab fixed frame to characterise the shear stress of atoms whose principal stress axis are randomly oriented with a magnitude $\hat{s}$ uncorrelated to the orientation.

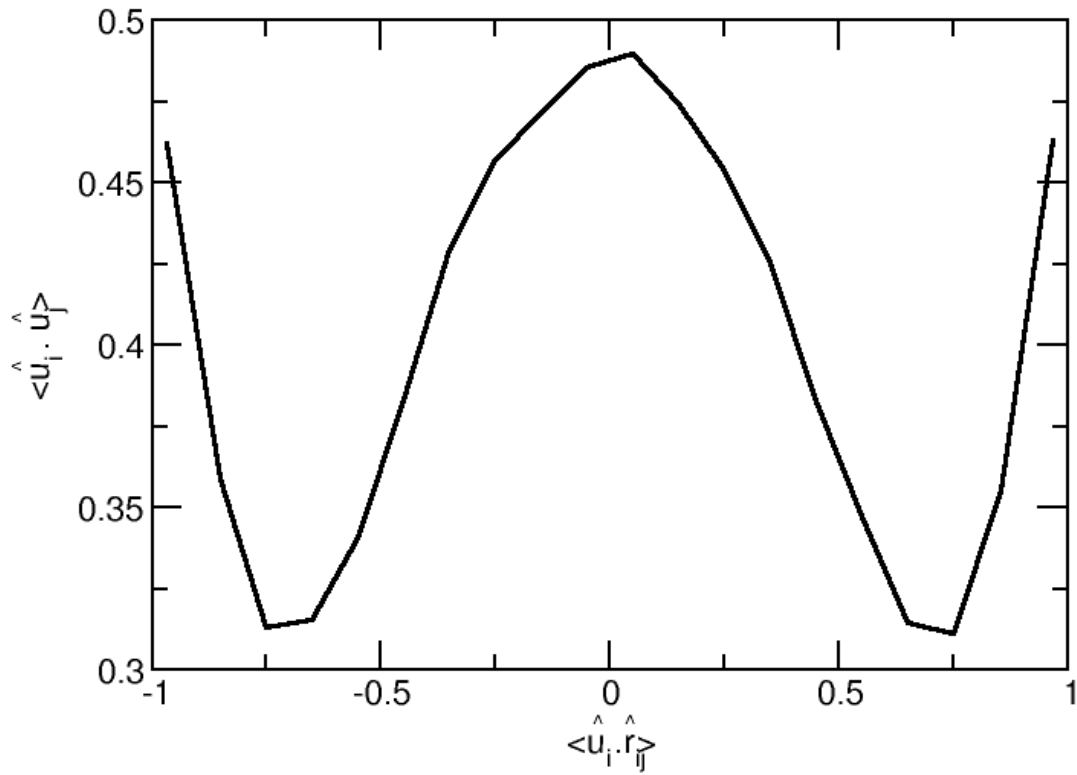

**Figure 11.** The mutual alignment $< \vec{u}_i \cdot \vec{u}_j >$ of the principal stress axes on adjacent atoms as a function of the cosine of the angle between $\vec{u}_i$ and the vector between particles i and j. The peaks at -1, 0 and 1 indicate the maximum alignment of the principal stress axes on adjacent atoms occurs when the two vectors are aligned collinearly. The data is for the same system as described in Fig. 10.

The $1/r^2$ decay of stress correlations along a given direction is consistent with the results of strain correlation in an elastic continuum [17]. On the level of the network of forces between neighbouring particles, we note that the local potential energy condition that requires that the net force on each particle vanishes results in a tendency of the principle stress axes to align into extended chains. To see this we have calculated $< \vec{u}_i \cdot \vec{u}_j >$, where particles i and j are neighbours, as a function of $< \vec{u}_i \cdot \hat{\vec{r}}_{ij} >$ where $\hat{\vec{r}}_{ij}$ is the unit displacement vector between the neighbours and find (see Fig. 11) that the alignment of the stress axes has maxima when the stress axes lie along $\hat{\vec{r}}_{ij}$, the condition for a chain-like correlation.

## 4. ON THE PHYSICAL SIGNIFICANCE OF INHERENT STRUCTURE STRESS

We have demonstrated here that the inherent structures exhibit the long range correlations characteristic of any elastic solid. In ref. 8 we argued that the IS stress was itself a manifestation of this correlation, arising as a consequence of the boundary conditions imposed in the simulation. Recently Fuereder and Ilg [21] have presented a derivation of the observed 1/N dependence of the IS stress variance in which the behaviour is explained as arising from the stress of atoms located near the edges of the simulation cell. This intrusion of the machinery of the simulation method into the physical fluctuations of the liquid deserves some comment. It is a basic notion of statistical mechanics that the choice of constraints determine the fluctuations one can observe. If we wish to study fluctuations in the shear stress then we must constrain the strain. It follows that the algorithmic machinery by which that constraint is applied (in this case, the fixed shape of the simulation cell) is an unavoidable feature of the calculation of the statistics of the stress fluctuations.

If we accept, then, that the IS stress and its statistical distribution is an intrinsic property of the energy landscape that a liquid explores, what physical aspects of the liquid does it influence? At high temperatures, the high mobility of particles renders the inherent structures irrelevant to the liquid properties. It is in the viscoelastic regime of supercooled liquids that the IS stress can influence the observed properties. In this regime, the stress relaxation function (a time dependent modulus) G(t) is given by [22],

$$G(t) = \beta V < \sigma(0)\sigma(t) > \qquad (6)$$

Eq. 6 is only valid when the time scale used to carry out the averages is long enough that zero frequency shear modulus $G_0 = 0$, the condition that we are considering a liquid. In ref. 8 it was suggested that the plateau modulus $G(t_p)$, where $t_p$ corresponds to a time on the plateau of the stress autocorrelation function $<\sigma(0)\sigma(t)>$, could be approximated by

$$G(t_p) \sim \beta V < \sigma_{IS}^2 > \qquad (7)$$

If the IS stress does indeed influence the plateau modulus and, hence, shear viscosity of the viscoelastic liquid then the long range correlations described here might give rise to some influence of the shape of the simulation cell. It is possible that these long range correlations are 'screened' by thermal fluctuations as discussed in ref. 18.

## 5. CONCLUSION

In conclusion, we have demonstrated that the atomic shear stress in the inherent structure of a quiescent liquid exhibits slowly decaying anisotropic correlations. We have shown that the anisotropy follows from the choice of the lab-fixed reference frame to define the shear stress. The slowly decaying character of this correlation reflects the correlations embedded in the force network in which forces must be balanced on all particles (i.e. the condition of a local potential minimum). We find that the shear stress of an inherent structure arises, in roughly equal parts, from a contribution of the atomic stress and their short range correlations and a size-independent long range contribution arising from the incomplete angular averaging in the 'corners' of the simulation cell. In contrast to previous authors [18-20] who attribute the anisotropic pattern of stress correlations to the elastic response to local rearrangements a la Eshelby [17], we have argued here that this pattern is the consequence of a random distribution of local principal stress axes projected onto the lab-fixed Cartesian reference frame and that the introduction of Eshelby 'events' is not required. What is clear from this and these previous studies, is that Frenkel's allusion [2] to the rigid state that underlies the liquid has real substance and that the interplay of the long range correlations of elasticity and the stress relaxation associated with viscosity in viscoelastic liquids remains to be fully understood.

**Acknowledgements**

We acknowledge support from the Australian Research Council.

**REFERENCES**

1. F. H. Stillinger, *Energy landscapes, inherent structures and condensed matter phenomena* (Princeton University Press, Princeton, 2015)

2. J. Frenkel, *The Kinetic Theory of Liquids* (Clarendon Press, Oxford, 1946)

3. F. Sciortino, W. Kob and P. Tartaglia, Phys. Rev. Lett. **83**, 3214 (1999); S. Sastry, Nature **409**, 164 (2001); R. Speedy, Mol. Phys**. 8**, 1105 (1993).

4. F. Sciortino, J. Stat. Mech.-Theory and Expt. **P05015** (2005)

5. A. Heuer, J. Phys.: Cond. Matt. **20**, 373101 (2008).

6. B. Doliwa and A. Heuer, Phys. Rev. E **67**, 031506 (2003); F.H.Stillinger, Science **267**, 1935 (1995).

7. J.-P. Hansen and I. R. Macdonald, *Theory of Simple Liquids* (Academic Press, London, 2006)

8. S. Abraham and P. Harrowell, J. Chem. Phys. **137**, 014506 (2012).

9. D. Perera and P. Harrowell, Phys. Rev. E **59**, 5721 (1999).

10. G. Wahnström, Phys. Rev. A **44**, 3752 (1991).

11. S. Toxvaerd, U. R. Pedersen, T. B. Schrøder and J. C. Dyre, J. Chem. Phys. **130**, 224501 (2009).

12. C. E. Maloney and A. Lemaître, Phys. Rev. E **74**, 016118 (2006).

13. C. Goldenberg, A. Tanguy and J.-L. Barrat. Europhys. Lett. **80**, 16003 (2007).

14. M. Tsamados, A. Tanguy, F. Léonforte and J.-L. Barrat, Eur. Phys. J. E **26**, 283 (2008).

15. E. Lerner and I. Procaccia, Phys. Rev. E **80**, 026128 (2009).

16. C. E. Maloney and M. O. Robbins, Phys. Rev. Lett. **102**, 225502 (2009).

17. J. D. Eshelby, Proc. Roy. Soc. (London) A **241**, 376 (1957).

18. B. Wu, T. Iwashita and T. Egami, Phys. Rev. E **91**, 032301 (2015).

19. J. Chattoraj and A. Lemaître, Phys. Rev. Lett. **111**, 066001 (2013).

20. A. Lemaître J. Chem. Phys. **143**, 164515 (2015).

21. I. Fuereder and P. Ilg, J. Chem. Phys. **142**, 144505 (2015).

22. H. Yoshino, J. Chem. Phys. **136**, 214108 (2012).